\documentclass[aps,prl,twocolumn,showpacs,amsmath,amssymb,amsfonts,nofootinbib,long,floatfix]{revtex4}
\usepackage{epsfig,latexsym,bm}

\newcommand\eV{\mbox{eV}}
\newcommand\GeV{\mbox{GeV}}

\newcommand\Mpc{\mbox{Mpc}}
\newcommand\G{\mbox{G}}
\newcommand\A{\mathbf{A}}
\newcommand\B{\mathbf{B}}
\newcommand\x{\mathbf{x}}
\newcommand\y{\mathbf{y}}
\newcommand\kk{\mathbf{k}}
\newcommand\ee{{\boldsymbol \varepsilon}}
\newcommand\mPl{m_{\rm Pl}}

\begin{document}

\title{Origin of Cosmic Magnetic Fields}

\author{Leonardo Campanelli$^{1}$}
\email{leonardo.campanelli@ba.infn.it}
\affiliation{$^1$Dipartimento di Fisica, Universit\`{a} di Bari, I-70126 Bari, Italy}

\date{\today}


\begin{abstract}
We calculate, in the free Maxwell theory, the renormalized quantum vacuum expectation value
of the two-point magnetic correlation function in de Sitter inflation. We find that quantum
magnetic fluctuations remain constant during inflation instead of being washed out
adiabatically, as usually assumed in the literature. The quantum-to-classical transition of
super-Hubble magnetic modes during inflation, allow us to treat the magnetic field classically
after reheating, when it is coupled to the primeval plasma. The actual magnetic field is
scale independent and has an intensity of $few \times 10^{-12} \G$ if the energy scale of
inflation is $few \times 10^{16} \GeV$. Such a field accounts for galactic and galaxy cluster
magnetic fields.
\end{abstract}




\maketitle


{\it Introduction.} -- The origin of the observed large-scale  $\mu \G$ magnetic fields in galaxies
and galaxy clusters is one of the major unsolved mysteries in cosmology
(for reviews on cosmic magnetic fields, see~\cite{Kronberg,Giovannini,Kandus et al}).

There are two main schools of thought about the generation of such cosmic magnetic fields,
according to which magnetic fields we observe today are created either in the early
Universe (``primordial hypothesis'') or during the processes of large-scale structure
formation and evolution (``astrophysical hypothesis'').
According to the primordial hypothesis, large-scale magnetic fields have been created
during an inflationary epoch of the Universe, or during primeval cosmic phase transitions
(such as electroweak or quark-hadron phase transitions). Successively, these relic fields
have been possibly amplified in galaxies and galaxy clusters by dynamo actions~\cite{Kronberg}.
The astrophysical hypothesis, instead, supposes that seed fields are generated by plasma
effects
directly in galaxies and galaxy clusters, and then amplified by a
dynamo mechanism. Both hypotheses meet with difficulties when their predictions are compared
with observations.

It is believed that inflation-produced magnetic fields have large correlation
scales $\lambda$ but extremely low intensities, unless some nonstandard physics is introduced, e.g.,
by adding nonstandard terms to the photon field Lagrangian~\cite{Kandus et al}.
As shown in~\cite{Barrow et al} (see~\cite{Shtanov-Sahni} for a recent criticism to this work),
this is the case only if the spatial curvature of
the Universe is zero.
However, in~\cite{Barrow et al}, the initial magnetic spectrum is that associated
to ``unrenormalized'' vacuum fluctuations. This is a questionable assumption,
since it gives a formally infinite, vacuum expectation value (VEV)
of the two-point magnetic correlation function.
It is the aim of this Letter
to bring into question the physical correctness of using unrenormalized
vacuum fluctuations and to show, contrary to what is believed, that strong
inflationary magnetic fields are a natural consequence of standard
quantum electrodynamics in curved space (in particular in a Friedmann spacetime
with zero spatial curvature). This is possible if one, in order to get a finite result,
``renormalizes'' the two-point magnetic correlator.

Phase-transition-generated fields can have astrophysically
relevant intensities~\cite{Giovannini},
but their correlation lengths are
too small
to explain cosmic magnetic fields, even allowing a possible amplification due to
magnetohydrodynamic turbulent effects operating in the early Universe~\cite{MHD}.

The generation of magnetic fields directly in galaxies and galaxy clusters is problematic
due to the fact that it is very difficult to explain
the presence of strong magnetic fields in galaxies at high redshift, since (large-scale) dynamo
actions are inefficient on short time scales~\cite{Kronberg}.
Moreover, the detected spectrum of distant
blazars~\cite{Neronov-Vovk}
seems to be compatible
with the presence of magnetic fields in voids, whose nature can be then explained only in the
framework of the primordial hypothesis.


{\it Seed fields.} -- The observation of magnetic fields in galaxies and
galaxy clusters could be explained if a sufficiently intense large-scale magnetic field,
such as
$10^{-13} \G \lesssim B_0 \lesssim few \times 10^{-12} \G$ with $\lambda \gtrsim few \times \Mpc$,
were present prior to their formation.
The above comoving values take into account the amplification and stretching of magnetic fields
inside galaxies and galaxy clusters, due essentially to the so-called Alfv\'{e}n frozen flux
effect~\cite{Giovannini} and
to the Kelvin-Helmholtz instability of
intracluster plasma flows~\cite{Dolag et al}.

In the following, we show that a primordial field with the above properties is
a natural consequence of inflation. To set notations and to explain why this kind of field
is believed not to be generated in the standard Maxwell theory, we consider first
the case analyzed in the literature, to wit, that of ``unrenormalized'' magnetic fluctuations
from inflation.


{\it Unrenormalized fluctuations.} --
The equation of motion for a magnetic field in a curved spacetime
is homogeneous in the field, so one needs an initial field in order to have a today field
different from zero. Quantum-mechanical effects during inflation give the unique possibility
to have such an initial magnetic field. As shown a long time ago by Parker~\cite{Parker-Toms},
particles can be created by quantum-gravitational effects in an expanding universe. However,
this is not the case for conformally invariant theories, a result known as ``Parker theorem.''
Standard electromagnetism in a Friedmann universe
is invariant under conformal transformations, so, in this case,
the only other way to have an initial magnetic spectrum is to consider electromagnetic vacuum fluctuations,
which are present even in conformally invariant theories.

The standard Maxwell Lagrangian for the electromagnetic
field $A_\mu$ is $\mathcal{L} = -\frac{1}{4} \sqrt{-g} F_{\mu \nu} F^{\mu \nu}$, where $g$ is
the determinant of the metric tensor and $F_{\mu \nu} = \partial_\mu A_\nu - \partial_\nu A_\mu$.
For the sake of simplicity, we assume that during inflation the Universe is described by a de
Sitter spacetime with line element $ds^2 = a^2(d\eta^2 - d \x^2)$, where $a$ is the expansion
parameter, $\eta = -1/(aH)$ is the conformal time, and $H$ is the (constant) Hubble parameter.
Working in the Lorentz gauge, $\nabla_\mu A^\mu = 0$, we expand the transverse part of the vector
potential as ${\A}_T(\x) = \sum_{\alpha=1}^2 \int \! d^3k \, (2\pi)^{-3} \ee_{\kk,\alpha} \,
a_{\kk,\alpha} \, A_{k,\alpha} \, e^{i\kk \x} + \mbox{H.c.}$, where the transverse polarization
vectors $\ee_{\kk,\alpha}$ satisfy the completeness relation
$\sum_\alpha (\varepsilon_{\kk,\alpha})_i (\varepsilon_{\kk,\alpha}^*)_j =
\delta_{ij}- k_i k_j /\kk^2$, with $\kk$ being the comoving wavenumber. The annihilation and
creation operators satisfy the usual commutation relations
$[a_{\kk,\alpha}, a_{\kk',\alpha'}^{\dag}] = (2\pi)^3 \delta_{\alpha \alpha'} \delta(\kk-\kk')$,
all the other commutators being null.

The equation of motion for $A_{k,\alpha}$ is $\ddot{A}_{k,\alpha} + k^2 A_{k,\alpha} = 0$
(a dot denotes differentiation with respect to the conformal time), whose solution is
$A_{k,\alpha} = c_1(k) \, e^{-ik\eta} + c_2(k) \, e^{ik\eta}$, with $c_1(k)$ and $c_2(k)$
constants of integrations. These are fixed by the choice of the vacuum, which is taken to
be the Bunch-Davies vacuum~\cite{Parker-Toms}.
In this case, the above constants are $c_1(k)=1/\sqrt{2k}$ and $c_2(k)=0$, so that we have
the standard plane-wave solution $A_{k,\alpha} = e^{-ik\eta}/\sqrt{2k}$.
Let us introduce the magnetic field, $\B(\x)$, in the usual way as $a^2 \B = \nabla \times \A_T$.
The vacuum expectation value of the squared magnetic field is then
$\langle 0| \B(\x)^2 |0 \rangle = \int_0^{\infty} dk k^{-1} \mathcal{P}(k)$, where
$\mathcal{P}(k) = \sum_\alpha [k^5/(2\pi^2 a^4)] |A_{k,\alpha}|^2$ is the so-called magnetic
power spectrum. For the plane-wave solution we have $\mathcal{P}(k) = k^4/(2\pi^2 a^4)$.

Introducing the comoving wavelength $\lambda$ as $k=2\pi/\lambda$, one usually defines the
magnetic field strength $B$ on the comoving scale $\lambda$ as
$B(\lambda) = \mathcal{P}(2\pi/\lambda)^{1/2}$. Accordingly, during de Sitter inflation the
magnetic field scales adiabatically, $B \propto a^{-2}$, reducing (exponentially) its intensity.
As a result, this field cannot explain cosmic magnetic fields, in agreement with the standard
literature~\cite{Turner-Widrow}.


{\it Quantum-to-classical transition.} -- Before analyzing the problem of
renormalization of inflationary quantum fluctuations, we notice that
a transition from quantum to classical behavior of
such fluctuations is generally expected to take place. Indeed, this occurs when quantum
coherence is destroyed by its coupling to the
environment. A quantum expectation value like $\langle 0| \B(\x)^2 |0 \rangle$ becomes then
indistinguishable from the corresponding classical ensemble average
$\langle \B(\x)^2 \rangle$~\cite{Polarski-Starobinsky}.

Classicalization of a given quantum fluctuation is realized when it crosses outside the horizon
during inflation, and this is understood in terms of its ``squeezing''
properties~\cite{Polarski-Starobinsky}. Once a given realization of the magnetic fluctuations
has occurred during inflation, further evolution proceeds classically. For this reason, we can
treat super-Hubble inflationary modes as classical stochastic fluctuations after inflation, and in
particular after reheating, namely after the energy associated to inflaton has been converted
into ordinary matter and any magnetic field get coupled to the newly formed plasma.


{\it Renormalized fluctuations.} -- The standard approach in calculating
the inflation-produced magnetic fluctuations is questionable since the quantity
$\langle 0| \B(\x)^2 |0 \rangle$ is formally infinite due to the ultraviolet divergence of the
power spectrum. This divergence can be cured by renormalizing the magnetic correlator. It is
worth noticing that the same situation appears in a very different context, namely in relation
to the primeval power spectrum of the cosmic microwave
background  radiation, when quantizing the inflaton field
fluctuations.
Here, renormalizing the inflaton
two-point correlator gives very significant effects on the amplitude and properties of
perturbations from inflation~\cite{Parker}.

In this Letter, we adopt the method of adiabatic renormalization~\cite{Parker-Fulling}
although, recently enough, there has been in the literature a critical discussion about
the validity of this renormalization technique~\cite{Adiabatic}.
In the adiabatic renormalization procedure, one assumes that the expansion parameter is a slowly varying function of time.
This is attained by replacing the expansion parameter $a(\eta)$ by a one parameter
family of functions $a_T(\eta) = a(\eta/T)$, and taking the limit of large ``slowness
parameter'' $T$.
This allows us to find a WKB (or adiabatic) solution to the equation of motions to any desiderate order
$(T^{-1})^n$ (with $n \geq 0$).
The adiabatic expansion is a formal one, in the sense that it must be applied even if $a(\eta)$ is not a slowly varying
function of time. This assures the conservation of the regularized energy-momentum tensor~\cite{Parker-Fulling}.
Then, the physical (i.e., renormalized) VEV of a given quantity
is obtained from the unrenormalized one by subtracting mode by mode
the corresponding adiabatic quantity up to the appropriate order,
the minimum adiabatic order being determined by the degree of
ultraviolet divergence of that quantity~\cite{Parker-Toms}.

We assume that the physical VEV is a linear operator, in the sense that
$\langle 0| \Psi_1[\psi(x)] + \Psi_2[\psi(x)] + ... |0\rangle_{\rm phys} =
\langle 0| \Psi_1[\psi(x)] |0\rangle_{\rm phys} + \langle 0| \Psi_2[\psi(x)] |0\rangle_{\rm phys} + ...$,
for all functions $\Psi_i$ of a given field $\psi$ evaluated at the spacetime point $x$.
This is a necessary condition we must impose on renormalized VEVs,
since this property is verified by classical ensemble averages and, according to the above discussion,
a possible classicalization of super-Hubble quantum fluctuations makes them indistinguishable from each other.
In order to cure ultraviolet divergences in the VEV of the energy-momentum tensor, $\langle 0|T^\mu_\nu |0\rangle$,
one generally needs to subtract from that, and mode by mode,
the corresponding adiabatic quantity up to the order $n=4$~\cite{Parker-Fulling}.
Since $T^\mu_\nu$ is constructed starting from local quadratic
quantities in the fields, the linearity of the $\langle 0| ... |0\rangle_{\rm phys}$ operator
requires the use of the fourth adiabatic order also for these quadratic quantities.
In order to renormalize the two-point magnetic correlator then, we
consider the WKB expansion up to fourth order.

In general, the adiabatic renormalization procedure applied to the stress tensor reduces to normal
ordering in the limit of static $a(\eta)$ (the Minkowski case),
and is completely equivalent to other renormalization schemes used in quantum theory
in curved spacetime~\cite{Birrell-Davies,Parker-Toms}.
In particular, it gives the correct value of the so-called ``conformal
anomaly'' in the case of conformally invariant theories~\cite{Birrell-Davies,Parker-Toms,Chimento-Cossarini}.
Moreover, in all renormalization schemes,
the removal of infinities in the VEV of the energy-momentum tensor
corresponds to the renormalization of the coupling constants in the Einstein's equations~\cite{Birrell-Davies}.

To apply the adiabatic renormalization procedure to the two-point magnetic correlator,
we firstly need to introduce a regulator photon mass, $m$, for the transverse part of the
vector potential, which will be sent to zero at the end of the calculation. This is possible
due to well-known fact that the transverse part of a Proca field (namely a massive spin-1
vector field) smoothly tends to the electromagnetic field in the limit of vanishing
mass~\cite{Deser}. This is also necessary since we must temporarily break conformal invariance of
electromagnetism otherwise the adiabatic solution and the exact solution to the equation of motion
would coincide. Generally, this would give incorrect results, such as
a vanishing electromagnetic conformal anomaly.
Similar breakdowns of conformal invariance happen in other
regularization schemes~\cite{Birrell-Davies}, such as dimensional regularization, where conformal invariance
is temporarily broken by letting the spacetime dimensions going away from $d=4$, or in the
$\zeta$-function regularization scheme, where the conformal invariance is broken by the
technique of analytic continuation.
(Analogue situations appear also in quantum theory in Minkowski spacetime. For example, dimensional
regularization may lead to the breaking of chiral invariance, giving a chiral anomaly.)
In the case of a massive photon, the equation of motion
becomes $\ddot{A}_{k,\alpha,m} + \omega^2 A_{k,\alpha,m} = 0$, where $\omega^2 = k^2 + m^2 a^2$.
The solution corresponding to the Bunch-Davies vacuum is
$A_{k,\alpha,m} = (\sqrt{\pi}/2) \, e^{i\pi(1+2\nu)/4} \sqrt{-\eta} \, H_\nu^{(1)}\!(-k\eta)$,
where $\nu = \sqrt{1/4-m^2/H^2}$ and $H_\nu^{(1)}\!(x)$ is the Hankel function of the first kind.

Second, we need the adiabatic solution $A_{k,\alpha,m}^{({\rm A})}$ of the equation of
motion. Because of the replacing of $a(\eta) \rightarrow a_T(\eta) = a(\eta/T)$ ($T \rightarrow \infty$),
the adiabatic order of the solution can be found by counting the number of time derivatives of
$a(\eta)$~\cite{Parker-Toms}.
Following the standard procedure~\cite{Parker-Toms}, we write
$A_{k,\alpha,m}^{({\rm A})} = e^{-i \! \int_0^\eta d\eta' W(k,\eta')}/\sqrt{2W(k,\eta)}$.
Expanding $W$ up to the fourth adiabatic order,
$W = \sum_{i=0}^4 \omega^{(i)}$,
we get, from the equation of motion,
$\omega^{(0)} = \omega$, $\omega^{(1)} = \omega^{(3)} = 0$,
$\omega^{(2)} = \frac{3}{8} \, \omega^{-3} \dot{\omega}^2 - \frac{1}{4} \, \omega^{-2} \ddot{\omega}$,
and $\omega^{(4)} = -\frac{297}{28} \, \omega^{-7} \dot{\omega}^4
- \frac{99}{32} \, \omega^{-6} \dot{\omega}^2 \ddot{\omega}
+ \frac{13}{32} \, \omega^{-5} \ddot{\omega}^2
+ \frac{5}{8} \, \omega^{-5} \dot{\omega} \dddot{\omega}
- \frac{1}{16} \, \omega^{-4} \ddddot{\omega}$.
The adiabatic expansion of $W^{-1}$
up to the fourth order, $W^{-1} =  \sum_{i=0}^4 (W^{-1})^{(i)}$, comes straightforwardly:
$(W^{-1})^{(0)} = \omega^{-1}$, $(W^{-1})^{(1)} = (W^{-1})^{(3)} = 0$,
$(W^{-1})^{(2)} = - \frac{3}{8} \, \omega^{-5} \dot{\omega}^2
+ \frac{1}{4} \, \omega^{-4} \ddot{\omega}$, and
$(W^{-1})^{(4)} = \frac{315}{28} \, \omega^{-9} \dot{\omega}^4
- \frac{105}{32} \, \omega^{-8} \dot{\omega}^2 \ddot{\omega}
+ \frac{15}{32} \, \omega^{-7} \ddot{\omega}^2
+ \frac{5}{8} \, \omega^{-7} \dot{\omega} \dddot{\omega}
- \frac{1}{16} \, \omega^{-6} \ddddot{\omega}$.

Finally, the physical
VEV of the squared magnetic field is
defined by the mode-by-mode (namely under the integral sign)
subtraction
$\langle 0| \B(\x)^2 |0\rangle_{\rm phys} =
\lim_{m \rightarrow 0} \! \int_0^{\infty} \! dk k^{-1} \mathcal{P}_{\rm phys}(k,m)$,
where we have defined $\mathcal{P}_{\rm phys}(k,m)
= \mathcal{P}(k,m) - \mathcal{P}^{({\rm A})}(k,m)$. Here, $\mathcal{P}(k,m) = \sum_\alpha [k^5/(2\pi^2 a^4)] |A_{k,\alpha,m}|^2$
is the exact magnetic power spectrum in the massive case,
while
$\mathcal{P}^{({\rm A})}(k,m) = \sum_\alpha \sum_{i=0}^4 [k^5/(2\pi^2 a^4)] (W^{-1})^{(i)}$
is the corresponding adiabatic expansion up to the fourth order.
We find that only the fourth-order term
determines the value of the renormalized magnetic correlator,
giving
$\langle 0| \B(\x)^2 |0\rangle_{\rm phys} = 19H^4/(160\pi^2)$~\cite{scalar}.
This shows that vacuum magnetic fluctuations during de Sitter inflation
are constant in time, and not adiabatically diluted by the cosmic expansion.

In order to study the correlation properties of these fluctuations, it is useful to
consider the two-point magnetic correlator. 
It can be expressed in terms of the power spectrum as
$\! \langle 0| \B(\x) \B(\y) |0\rangle =
\int_0^{\infty} \! dk k^{-1} \mathcal{P}(k) j_0\!(k|\x-\y|)$, where $j_0\!(x)$ is the
zeroth-order spherical Bessel function of the first kind. The physical two-point magnetic
correlation function is, adopting again the adiabatic renormalization scheme,
$\langle 0| \B(\x) \B(\y) |0\rangle_{\rm phys} = \lim_{m \rightarrow 0} \!
\int_0^{\infty} \! dk k^{-1} \mathcal{P}_{\rm phys}(k,m) j_0\!(k|\x-\y|)$,
giving
\begin{equation}
\label{Correlator}
\langle 0| \B(\x) \B(\y) |0\rangle_{\rm phys} = \frac{19H^4}{160\pi^2} \, .
\end{equation}
This implies that $\mathcal{P}_{\rm phys}(k,m)/k$, the double of the so-called magnetic
energy density spectrum, is asymptotically proportional to a delta function, $\delta(k)$,
in the limit $m \rightarrow 0$. Physically and in contrast to the case of unrenormalized
fluctuations, this means that magnetic vacuum fluctuations do not depend on the comoving
scale $\lambda = |\x-\y|$~\cite{scale,zeta}. Thereby, inflation ``grows'' quantum fluctuations
equally on sub- and superhorizon scales~\cite{Hubble}.


{\it Backreaction on inflation.} -- The above calculations have been carried out in a fixed
de Sitter background,
namely assuming that backreaction of electromagnetic
vacuum fluctuations on inflation is negligible.
This is valid if the physical VEVs of
the components of the electromagnetic energy-momentum tensor are much smaller than those associated to inflation,
$(T^\mu_\nu)_{\rm inf} = M^4 \delta^\mu_\nu$, where
$\delta^\mu_\nu$ is the Kronecker delta. Here, we have introduced the energy scale of inflation,
$M$, which is related to the energy density of inflation, $\rho_{\rm inf}$, through
$M^4 = \rho_{\rm inf} = 3H^2/(8\pi G)$, where $G=1/\mPl^2$ is the Newton constant
and $\mPl$ is the Planck mass.

The physical VEV of the electromagnetic energy-momentum tensor cannot be obtained as the massless
limit of the total (transverse plus longitudinal) energy-momentum tensor of the Proca field.
This is due to the fact that the longitudinal part of the energy-momentum tensor in the massive theory
is not well behaved as $m \rightarrow 0$~\cite{Chimento-Cossarini}.
In this case, to get the right result one needs to add
a gauge-breaking term and a compensating complex ghost field to the standard Proca
Lagrangian~\cite{Chimento-Cossarini}. The final result is the usual one,
$\langle 0|(T^\mu_\nu)_{\rm e.m.}|0 \rangle_{\rm phys} = (31/480\pi^2) H^4 \delta^\mu_\nu$~\cite{Parker-Toms},
and is strictly connected to the electromagnetic conformal
anomaly.
Consequently, backreaction on inflation is negligible if $(M/\mPl)^4 \ll 135/62$, which essentially
means that the energy scale of inflation must be below the Planck scale
$\mPl \simeq 1.22 \times 10^{19} \GeV$.


{\it The renormalized actual field.} -- To simplify the analysis we consider the case
of instantaneous reheating; i.e., we assume that after inflation the
Universe enters directly in the radiation dominated era.
From the beginning of this era
till the present time, quantum magnetic vacuum fluctuations are decohered and can be treated
as classical stochastic fluctuations. In the presence of a plasma with conductivity $\sigma$,
a classical magnetic field evolves according to the autoinduction equation~\cite{Turner-Widrow}
$\partial{(a^2 \B)/\partial \eta} = (1/\sigma)\nabla^2 (a^2 \B)$. In the limit of (infinitely)
high conductivity, we get $a^2 \B (\x,\eta) = a^2_{\rm RH} \B (\x,\eta_{\rm RH})$, where RH
indicates the time of reheating.
Accordingly, we have
$\langle \B(\x,\eta) \B(\y,\eta) \rangle =
\langle \B(\x,\eta_{\rm RH}) \B(\y,\eta_{\rm RH})\rangle (a_{\rm RH}/a)^4$,
where the classical ensemble average
$\langle \B(\x,\eta_{\rm RH}) \B(\y,\eta_{\rm RH}) \rangle$
is indistinguishable from the quantum correlator $\langle 0| \B(\x) \B(\y) |0\rangle_{\rm phys}$
on large (super-Hubble) scales, as explained above.

Since $a \propto g_{*S}^{-1/3} T^{-1}$ after reheating, where $g_{*S}(T)$ is the effective
number of entropy degrees of freedom at the temperature $T$~\cite{Kolb-Turner}, the actual
value of the magnetic field intensity is
$B_0 = B_i (\, g_{*S,0}/g_{*S,\rm{RH}})^{2/3} (T_0/T_{\rm RH})^2 \cos \theta_{W}$.
Here, $B_i$ is the root-mean-square value of the physical magnetic field at the end of
inflation, $T_0 \simeq 2.37 \times 10^{-4} \eV$ is the actual temperature, $T_{\rm RH}$ is
the reheat temperature,
$g_{*S,0} = g_{*S}(T_0) = 43/11$~\cite{Kolb-Turner},
and $g_{*S,\rm{RH}} = g_{*S}(T_{\rm RH})$. Above the electroweak phase transition (when we
assume inflation is taking place) the $U(1)$ gauge field which is quantum mechanically
excited is indeed the hypercharge field, not the electromagnetic one. Below the electroweak
phase transition, however, the hypercharge field is projected onto the electromagnetic field,
and this gives the cosine of the Weinberg angle $\theta_{W}$.

The reheat temperature can be related to the energy scale of inflation by observing that the
energy density of radiation at the beginning of radiation era,
$\rho_{\rm rad} = (\pi^2/30) g_{*,\rm{RH}} \, T_{\rm RH}^4$,
where $g_{*,\rm{RH}}$ is the effective number of degrees of freedom at the time of reheating
and can be taken equal to $g_{*S,\rm{RH}}$~\cite{Kolb-Turner}, must be equal to the energy
density at the end of inflation. We get $T_{\rm RH} = [30/(\pi^2 g_{*,\rm{RH}})]^{1/4} M$.
Taking $g_{*S,\rm{RH}} = 427/4$~\cite{Kolb-Turner}, referring to the massless degrees of
freedom of the standard model of particle physics, 
the actual, scale-independent magnetic field is
$B_0 \simeq 3 \times 10^{-13} (M/10^{16} \GeV)^{2} \G.$
If the energy scale of inflation is around $M \simeq 10^{16} \GeV$,
this field explains the cosmic magnetic fields.


{\it Conclusions.} -- We have shown, in the framework of the standard free Maxwell theory,
that the renormalized quantum VEV of the two-point magnetic correlation function does not
evolve adiabatically but remains constant during de Sitter inflation. Quantum magnetic
fluctuations are scale independent and their intensity depends on the scale of inflation.
Super-Hubble quantum magnetic fluctuations decohere during inflation, and can be then
treated as classical stochastic fluctuations in radiation and matter eras, when they are
coupled to the cosmic plasma. The actual magnetic field is scale independent on large scales
and, if the scale of inflation is of order of $M \sim 10^{16} \GeV$, it has the right intensity to
explain the magnetization of galaxies and galaxy clusters.


\begin{acknowledgments}
We would like to thank
G.~Florio, A.~Marrone, and L.~Parker for useful discussions.
\end{acknowledgments}




\begin{thebibliography}{99}
\bibitem{scalar}               As a check of the above results, we can calculate the transverse part of the
                               renormalized energy-momentum tensor for the Proca field and then take the limit
                               $m \rightarrow 0$. We straightforwardly obtain that it is twice the renormalized
                               energy-momentum tensor for a massless conformally coupled scalar field, as it
                               should be~\cite{Chimento-Cossarini}.
\bibitem{zeta}                 The fact that the physical magnetic VEV is (scale-) and time-independent,
                               should not come as a surprise. Indeed, the same is true for the
                               renormalized electromagnetic energy density (see below).
                               Moreover, this result is also confirmed in the $\zeta$-function
                               renormalization approach. To see this, we observe that
                               $\langle 0| \B(\x)^2 |0\rangle \sim \int_0^{\infty} \! dk k^3/a^4 \sim
                               H^4 \!\int_0^{\infty} \! dx x^3 \sim H^4 \sum_{n=1}^{\infty} n^3$,
                               where we introduced the dimensionless variables $x = -k\eta$, and roughly
                               approximated the integral by a sum. The divergent sum can now be cured by
                               analytical continuation~\cite{Birrell-Davies} by replacing it with
                               $\zeta(-3) = 1/120$, where $\zeta(x)$ is the Riemann $\zeta$-function.
                               This gives the renormalized and finite result
                               $\langle 0| \B(\x)^2 |0\rangle_{\rm phys} \sim H^4$.
\bibitem{Hubble}               We expect that the results obtained in de Sitter inflation hold,
                               with minor modifications, also in more realistic models of inflation.
                               This because while the Hubble parameter sets the scale of energy during inflation,
                               giving $\langle 0| \B(\x)^2 |0\rangle_{\rm phys} \sim H^4$, the mass of the field sets
                               the scale of correlation. For the photon case, then, the
                               renormalized vacuum magnetic fluctuations are allowed to be correlated
                               on arbitrarily large distances.
\end{thebibliography}
\end{document}